\documentclass[aps,prl,twocolumn,groupedaddress]{revtex4-2}
\usepackage{aas_macros}
\usepackage{tikz}
\usepackage{amsmath}
\usepackage{bm}
\usepackage{makecell}
\DeclareMathOperator{\sech}{sech}
\usepackage{xcolor}

\begin{document}


\title{
Efficient Nonthermal Ion and Electron Acceleration Enabled by the
Flux-Rope Kink Instability in 3D Nonrelativistic Magnetic Reconnection}


\author{Qile Zhang}
\affiliation{Los Alamos National Laboratory, Los Alamos, NM 87545, USA}
\email[]{qlzhanggo@gmail.com}

\author{Fan Guo}
\affiliation{Los Alamos National Laboratory, Los Alamos, NM 87545, USA}
\author{William Daughton}
\affiliation{Los Alamos National Laboratory, Los Alamos, NM 87545, USA}
\author{Xiaocan Li}
\affiliation{Dartmouth College, Hanover, NH 03755, USA}
\author{Hui Li}
\affiliation{Los Alamos National Laboratory, Los Alamos, NM 87545, USA}


\date{\today}

\begin{abstract}
The relaxation of field-line tension during magnetic reconnection gives rise to a universal Fermi acceleration process involving the curvature drift of particles.  However, the efficiency of this mechanism is limited by the trapping of energetic particles within flux-ropes.  Using 3D fully kinetic simulations, we demonstrate that the flux-rope kink instability leads to {\color{black}strong} field-line chaos in weak-guide-field regimes where the Fermi mechanism is most efficient, thus allowing particles to transport out of flux-ropes and undergo further acceleration.  As a consequence, both ions and electrons develop clear power-law energy spectra which contain a significant fraction of the released energy. The low-energy bounds are determined by the injection physics, while the high-energy cutoffs are limited only by the system size. These results have strong relevance to observations of nonthermal particle acceleration in space and astrophysics.
\end{abstract}


\maketitle


\textit{Introduction.---}
Within space and astrophysical plasmas, magnetic fields often develops stressed current sheets which are susceptible to magnetic reconnection -- a process that rapidly reconfigures the magnetic topology leading to high-speed flows, plasma heating, and nonthermal particle acceleration \citep{Yamada2010}.   Understanding this acceleration physics has immediate applications to the magnetosphere and solar flares, and various astrophysical problems. Observations from both the solar corona \citep{Lin2011,Omodei2018,Gary2018,Shih2009} and magnetotail \citep{Ergun2018,Ergun2020} show simultaneous productions of ion and electron power-law energy distributions extending to high energy during reconnection, suggesting a common physical origin.    However, the underlying physics remains poorly understood, since researchers have previously failed to produce these power-laws within self-consistent kinetic simulations in the relevant regime. 

On the theoretical front, previous studies have demonstrated a Fermi-type mechanism within reconnection layers \citep{Drake2006,Dahlin2014,Dahlin2017,Li2017,Li2018,Li2019b}, involving the particle curvature drift within the electric field induced by the large-scale flows.   
This mechanism driven by field-line curvature is efficient for low guide field (out of the reconnection plane) to reconnection field ratios $b_g<0.5$ 
\citep{Li2017,Li2018,Li2019b,Arnold2021}, and is enhanced by magnetic-island interactions within the reconnection layer.  However, in 2D simulations, the efficiency of the Fermi acceleration is limited by particle trapping within these islands.   Several 3D studies have demonstrated that overlapping tearing islands due to multiple resonance surfaces lead to field-line chaos \citep{Bowers2007,Daughton2011,Liu2013prl,Onofri2006}, allowing energetic particles to transport out of flux-ropes and continue acceleration \citep{Dahlin2015,Dahlin2017,Li2019b}.   However, this mechanism only applies to regimes with significant guide fields ($b_g>0.5$) where the efficiency of the Fermi mechanism is dramatically reduced.

With 3D kinetic simulations, here we demonstrate  that within the weak-guide-field regime the flux-rope kink instability is unstable in the reconnection layer. 
This generates strong field-line chaos, allowing energetic particles to transport out of flux-ropes and continue Fermi acceleration.
The field-line chaos is triggered when flux-ropes reach a threshold length for the $m=1$ kink instability. 
 Our 3D simulations exploiting this threshold reach an unprecedented domain size.
For the first time, both protons and electrons develop clear and sustainable nonthermal power-laws. The nonthermal populations contain a significant fraction of the released energy, and nonthermal protons gain $\sim2\times$ more energy than nonthermal electrons. The acceleration processes include an injection followed by a prolonged Fermi-acceleration phase. While the injection sets low-energy bounds of the power-laws which controls the nonthermal energy contents, the high-energy cutoffs keep growing with system size, indicating that the results can be extended to macroscopic systems. 


\textit{Numerical Simulations.---\label{setup}}
We use VPIC code that solves the Vlasov-Maxwell equations \citep{Bowers2008}. The 3D simulations start from a force-free layer $\mathbf{B}=B_0 \tanh(z/\lambda)\mathbf{e_x}+\sqrt{B_0^2 \sech^2(z/\lambda)+B_g^2}\mathbf{e_y}$ with a uniform plasma density $n_i = n_e = n_0$. $B_0$ is the reconnecting field, $B_g$ is the guide field and $\lambda$ is the half-thickness of the layer, which is set to be one ion inertial length $d_i$. Electrons carry the initial current that satisfies the Ampere's law. Most simulations have proton-to-electron mass ratio $m_i/m_e=25$, $b_g=B_g/B_0=0.2$ and $V_A=B_0/\sqrt{4\pi n_0m_i}=0.2c$, where $c$ is the speed of light. The initial temperature $T_i=T_e=0.01m_iV_A^2$ so the plasma $\beta$ based on the reconnecting field $\beta=0.02$. The grid size is $\Delta x=\Delta y=\Delta z=0.0488d_i$, with 150 particles per cell per species. Boundary conditions are periodic in $x$ and $y$, and conducting for fields  and reflecting for particles in $z$. A small long-wavelength perturbation is included to initiate reconnection. To limit the influence of periodic boundaries, all simulations terminate around $1.3$ Alfv\'en crossing time $L_x/V_A$ before the acceleration stagnates. {During this time, $\sim1/3$ of the upstream flux is reconnected and thus the influence of the $z$ boundary condition is minimal. } {\color{black} Our simulations are important for multi-X-line collisionless reconnection, and also relevant for a hierarchy of collisional plasmoids that may develop kinetic-scale current layers to trigger collisionless reconnection \citep{Shibata2001,Comisso2017,Daughton2009,Ji2011}.} A set of simulations have been conducted to confirm the robustness of the underlying processes for different guide fields, $\beta$, domain sizes, and mass ratios. See Supplemental Material for a summary. 

\begin{figure*}
\includegraphics[width=0.95\textwidth]{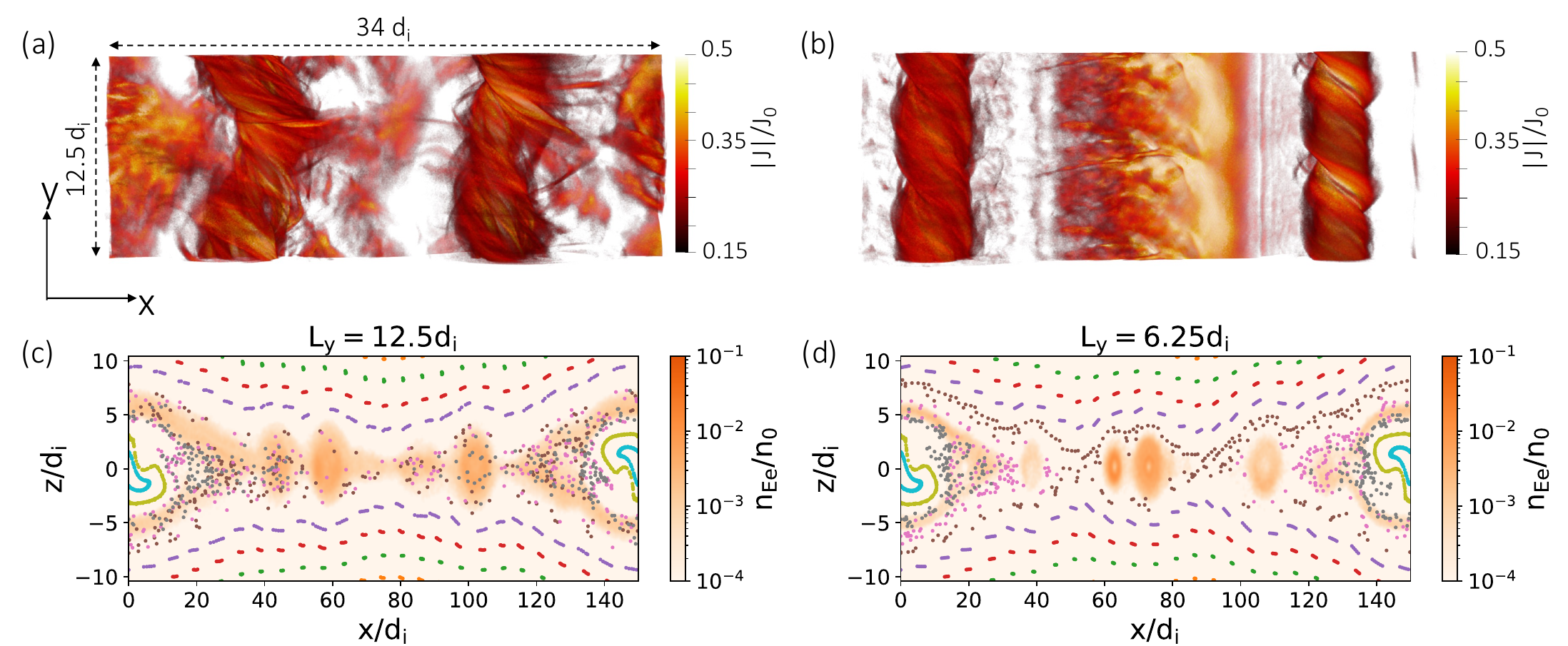}
\caption{Current density for simulations with $L_x=150d_i$ and different y-dimensions (a) $L_y=12.5d_i$ and (b) $L_y=6.25d_i$, respectively. 
(c) and (d) show respectively the $y$-averaged energetic electron density in these two simulations, overplotted with Poincar\'e-type plots of magnetic field lines -- traced from $x=0$ with their locations in the $x-z$ plane recorded every $6.25 d_i$ in $y$, and with different colors for different starting points. \label{fig1}}
\end{figure*}

\textit{Kink Instability and Threshold for 3D Effects.---}
Figure \ref{fig1}(a) shows the current density of flux-ropes at $\Omega_{ci}t=100$ in a simulation with $L_x \times L_y \times L_z =150 \times 12.5 \times 62.5d_i^3$ \footnote{The rendering has used a lower limit of $|J|$ to filter the upstream contents and emphasize the central regions of flux-ropes to better visualize their motions.}. The flux-ropes undergo $m=1$ kink instability and its nonlinear evolution tears up the flux surfaces (see also the supplemental movie). For comparison, flux-ropes in the simulation with $L_y = 6.25 d_i$ (same $L_x$ and $L_z$) do not have such dynamics (Figure \ref{fig1}(b) \footnote{The domain in Figure \ref{fig1}(b) has been replicated in $y$ as it is periodic, to better compare with Figure \ref{fig1}(a).}), although high-harmonic kink modes may develop. Figure \ref{fig1}(c) and (d) show the $y$-averaged energetic electron density ($1.2 < \varepsilon/m_iV_A^2 < 2.4$) of these two simulations, overplotted with Poincar\'e-type plots of magnetic field lines. Figure \ref{fig1}(c) shows that the kink instability drives strong field-line chaos mixing up different flux surfaces while Figure \ref{fig1}(d), in contrast, is nearly laminar. 

The transition into the {\color{black} strong} field-line chaos occurs when $L_y$ reached a threshold $L_{th}$, controlled by the criterion of $m=1$ kink disruption, namely the safety factor at the edge of flux-ropes $q_c =\pi b_gD/L_{th} \sim 1$ \citep{Oz2011}, where $D$ is the flux-rope diameter. 
Closer examination finds $D\sim 15d_i$ and $L_{th} \sim 9.5d_i$, placing the $L_y=12.5d_i$ case above the threshold and the other one below it, consistent with the observed dynamics. 
{\color{black}{This is in contrast to the earlier considered overlapping oblique tearing modes in the strong-guide-field regime~\cite{Bowers2007,Daughton2011,Liu2013prl}}}. In Supplemental Material, we systematically verified this threshold and field-line chaos driven by the kink instability in the low-guide-field regime ($b_g<0.5$). 

\begin{figure*}
\includegraphics[width=\textwidth]{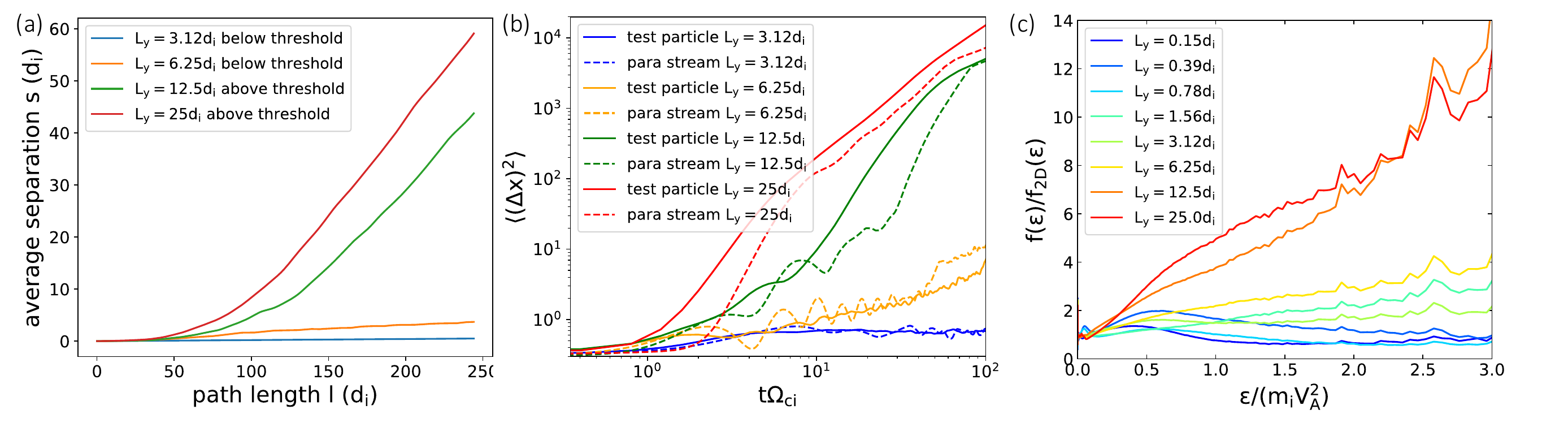}
\caption{(a) the averaged separation of initially adjacent field lines in the $x-z$ plane traced from the center of flux-ropes with $D\sim15d_i$.
(b) the mean-square displacement in $x$ for test-particle electrons and parallel-streaming particles traced from the cores of the flux-ropes.
Test-particle electrons are injected with an initial isotropic velocity $\sim3.5V_A$ while the parallel-streaming particles have a parallel velocity equal to the test-particles' root-mean-square parallel velocity.
(c) the enhancement of energetic electrons in 3D compared with a 2D simulation. \label{fig2}}
\end{figure*}

The field-line chaos leads to particle transport out of flux-ropes and further acceleration in the reconnection layer. 
Figure \ref{fig1}(c) also shows energetic electrons spreading out of flux-ropes, whereas in Figure \ref{fig1}(d) the electrons are effectively trapped in flux-ropes. 
The chaotic fields can be further understood by field-line separations, namely the distances between field-line pairs with small initial displacements \citep{Yang2020,Guo_2020}.
Figure \ref{fig2}(a) shows the averaged separation of $10^3$ field-line pairs, starting from the center of flux-ropes with $D \sim 15d_i$ (e.g., $x\sim58d_i$ in Figure \ref{fig1}(c) and $x\sim73d_i$ in Figure \ref{fig1}(d)) for several simulations with different $L_y$ (the same $L_x$ and $L_z$  as Figure \ref{fig1}). Above-threshold cases are far more chaotic (faster separation) than below-threshold cases. To better quantify particle transport, we also trace test-particle electrons with an isotropic initial velocity $V \sim 3.5V_A$ from centers of the flux-ropes (Figure \ref{fig2}(b)). Above-threshold cases consistently show stronger transport leaving the center of flux-ropes.
To understand the transport mechanism, we calculate the displacement assuming particles just stream along field lines with a parallel speed $V_\parallel = 2 V_A$ (root-mean-square value of the test-particle parallel velocities), which shows a trend similar to the test-particles (Figure \ref{fig2}(b)). This suggests that streaming along the chaotic field lines is an important mechanism for particles to transport out of flux-ropes.
Particle transport enabled by flux-rope kink instability greatly enhances the efficiency of particle acceleration (Figure \ref{fig2}(c)). Above-threshold cases consistently produce about $\sim10$ times more energetic particles (at energies $\sim100$ times of the initial thermal energy) than the 2D case, whereas below-threshold cases only show moderate increase.  

\textit{Simultaneous Nonthermal Acceleration of Protons and Electrons.---}
For larger reconnection domains, $L_{th}\sim\pi b_g D\sim 0.1\pi b_g L_x$ as $D$ grows with $L_x$, verified by simulations with different sizes. Our 3D simulations exploiting this condition extend to an unprecedented reconnection domain ($L_x \times L_y \times L_z=300 \times 25 \times 125d_i^3$). We discuss nonthermal acceleration revealed by this simulation.

\begin{figure*}
\includegraphics[width=0.75\textwidth]{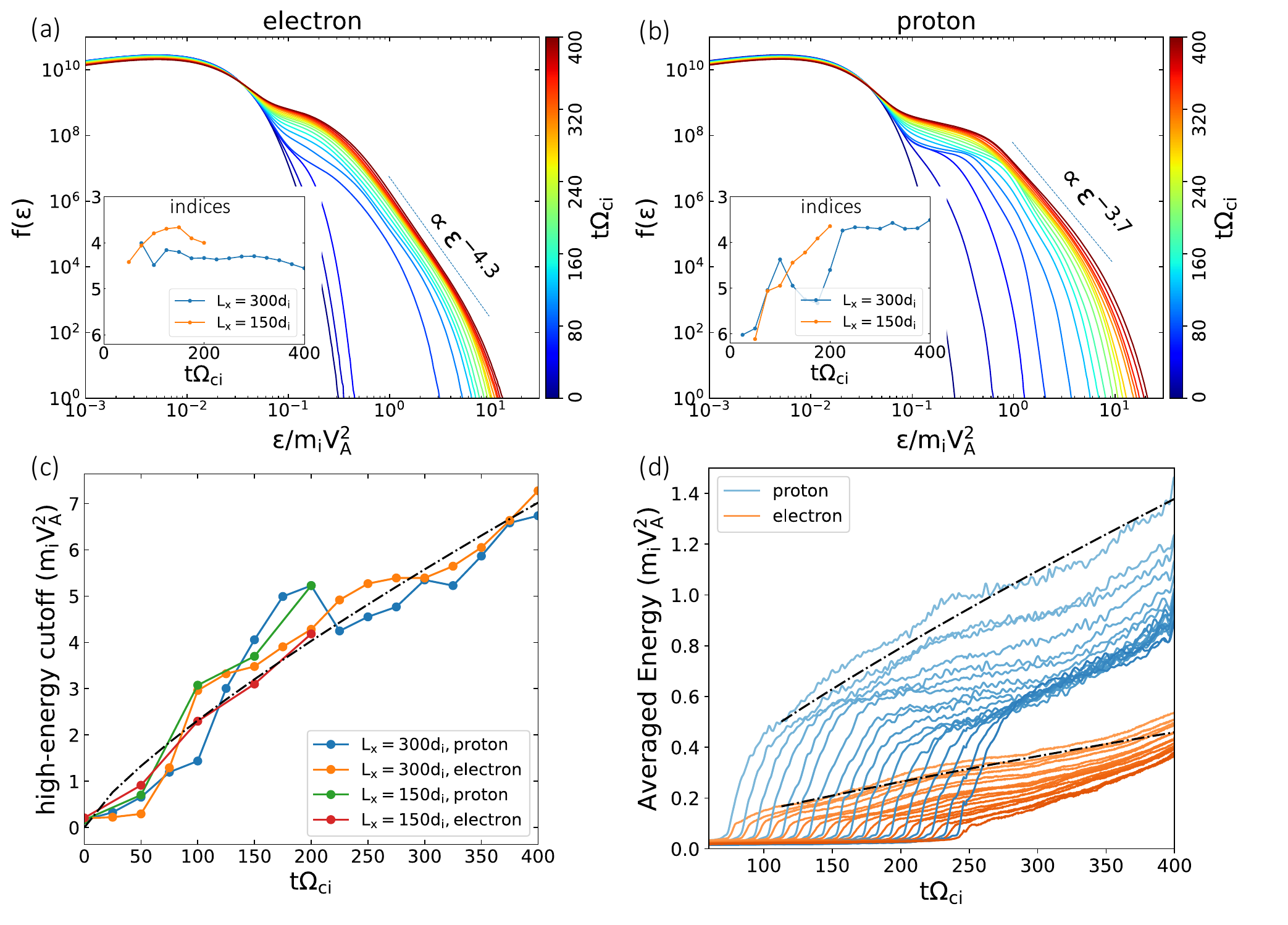}
\caption{Evolution of energy spectra for electrons (a) and protons (b) in the $L_x=300 d_i$ simulation. The spectral indices of this and a $L_x=150 d_i$ simulation are shown in the insets. (c) shows the high-energy cutoff of the power-laws, determined by the energy at which the spectrum deviates from the fitted power-law by 50\%. (d) shows the energization history of different generations of injected particles. The dash-dot lines represent the $\varepsilon\propto t^{0.8}$ scaling in (c) and (d).\label{fig3}}
\end{figure*}

Figure \ref{fig3}(a) and (b) show time evolution of energy spectra {\color{black} over the whole domain} for electrons and protons with insets showing the corresponding spectral indices. Both electrons and protons evolve into clear power-laws. While a smaller simulation ($L_x = 150d_i$) shows variable indices, the largest simulation shows that both electrons and protons sustain steady indices. Interestingly, protons took longer to settle into a steady power-law ($\Omega_{ci}t \sim 225$), which has been challenging to achieve in previous simulations. {\color{black}Due to our simulation parameters, high-energy electrons become mildly relativistic, making their spectra softer \citep{Caprioli2014a, Haggerty2019}. More analysis shows protons and electrons have the same spectral index in momentum spectra \citep{Guo2016}}. The low-energy bounds of the power-laws are nearly constants $\varepsilon_{le}\sim 0.2 m_i V_A^2$ (electron) and $\varepsilon_{li}\sim0.5 m_i V_A^2$ (proton) over time. Meanwhile, the high-energy cutoffs persistently increase with longer time and larger domains (Figure \ref{fig3}(c)), reaching $\sim 500$ times of the initial thermal energy. This suggests the nonthermal spectra can extend to much higher energies in macroscopic systems.

The first-ever proton and electron power-laws in kinetic simulations
 reveal the properties of the nonthermal components in magnetic reconnection. We determine the reconnected population by subtracting the cold upstream thermal distribution from the {\color{black}whole} distribution. Out of this population, $\sim20\%$ of particles and $\sim50\%$ of energy for each species are nonthermals above the low-energy bounds.  This efficiency is consistent with observations during solar flare reconnection \citep{Krucker2010,Oka2015,Aschwanden_2016,Arnold2021}. Energetic protons gain about twice as much energy as electrons, likely due to a more efficient injection process, as we will discuss later. 
 
In contrast, the spectra for below-threshold cases do not form clear and sustainable power-laws for either species. Therefore, $L_y>L_{th}$ also serves as a condition for power-law formation. We have verified the dependence of $L_{th}$ on guide fields using simulations with $b_g<0.5$ and {\color{black}that the electron and proton spectra are insensitive to mass ratios in the range $25-100$.}

\textit{Particle Acceleration Processes.---} 
Figure \ref{fig3}(d) shows the energy evolution of different generations of ions and electrons. Particles are selected if their final energies are above the corresponding low-energy bound, and are averaged as one ``generation'' if the starting time of energization is within a $\Omega_{ci}\Delta t=5$ interval. The energy evolution of each generation suggests that the accelerated particles experience an injection process followed by a prolonged Fermi-acceleration phase{\color{black}, analogous to the two-stage acceleration process in relativistic turbulence \citep{Comisso2018,Comisso2019}.}
The low-energy bounds of the power-laws are determined by the injection energy. When cold protons first {\color{black}cross a} reconnection exhaust, they receive an initial kick from the outflow ($V_{out}\sim0.5V_A$ in our simulations{\color{black}, due to low $b_g$ \citep{Haggerty2018,Li2021}}) and gain a speed of about $2V_{out}$ (with $\varepsilon \sim \varepsilon_{li}$), which boosts their energy for further acceleration. Thus at early time (e.g., $\Omega_{ci}t=75$), most protons are bounded by $\varepsilon_{li}$ in Figure \ref{fig3}(b), and later acceleration shapes the distribution into a power-law extending to higher energy. We have confirmed $\varepsilon_{li}\sim0.5m_iV_A^2$ with different simulations of low-$\beta$ ($\lesssim 0.1$).
On the other hand, electrons are much lighter, resulting in a less efficient energization process in the exhaust{\color{black}, where parallel electric fields could be important~\citep{Phan2013,Shay2014,Zhang2019,Zhang2019b,Le2009,Haggerty2015,Dahlin2017,Comisso2018}}. As a result, the electrons' low-energy bound $\varepsilon_{le} \sim 4 T_{ex}$, where $T_{ex}$ is the electron temperature at the exhaust ($\sim0.05m_iV_A^2$ in this simulation). 
{Note that $ T_{ex}$ can depend on many parameters ($\beta$, $B_g$, etc.), which will be studied in the future. }

The Fermi acceleration process can be elucidated by the following scaling analysis. 
Since the Fermi acceleration rate at typical acceleration regions (exhausts) $\alpha\equiv{\dot{\varepsilon}}/\varepsilon\sim \mathbf{U_E}\cdot\bm{\kappa}\sim V_{Ax}\kappa_x$ and the escape rate from the reconnection layer $\tau_{esc}^{-1}\sim V_{Ax}/L$, the power-law index \citep{Drury1999, Guo_2014}
\begin{equation}
p=1+(\alpha \tau_{esc})^{-1}
 \sim 1+\frac{V_{Ax}/L}{V_{Ax}\kappa_x}
 =1+\frac{1}{L \kappa_x},
\end{equation}
where $\mathbf{U_E}$ is the E$\times$B drift speed, $\bm{\kappa}$ is the magnetic field curvature vector and L is the half length of the reconnecting current sheet. Using $\kappa_x=(\hat{b}\cdot\nabla\hat{b})_x\sim \hat{b}_z\partial_z \hat{b}_x
\sim B_z B_x/(B^2\Delta_z$), 
where $\Delta_z$ is the typical length scale of exhaust field lines in z (related to the scale of flux-ropes), we obtain

\begin{equation}
p\sim 1+\frac{B_x}{B_z}\frac{\Delta_z}{L}(1+\frac{B_g^2}{B_x^2}).
\end{equation}
\begin{equation}
\alpha\sim V_{Ax}\kappa_x=\frac{B_z V_{Ax}B_x^2}{B_x(B_x^2+B_g^2)\Delta_z}.
\end{equation}
Since both $\Delta_z$ and $L$ are proportional to the domain size, in larger simulations the predicted spectral indices remain the same.
In the acceleration regions (exhausts), taking $B_g^2/B_x^2 \ll 1$, $B_z\sim0.05B_x$, $\Delta_z/L\sim0.15$ typical during our simulations, we obtain $p\sim4$, which is comparable to indices in our simulations. 
Since flux-ropes grow over time,  $\Delta_z$ increases (approximately linearly) and leads to a decrease in the acceleration rate
$\alpha\sim C/t$, where C is a constant. More careful inspection to the time evolution of $\Delta_z$ suggests $C\sim0.8$ (not shown). We also measure the acceleration rates directly from simulations as in Li et al. \citep{Li2019b}, finding values and dependence on time and simulation size consistent with the theoretical prediction. From the acceleration rate above, we obtain particle energy $\varepsilon\propto t^{0.8}$. This scaling agrees reasonably well with the growth of high-energy cutoffs (Figure \ref{fig3}(c)) and particle energy evolution (Figure \ref{fig3}(d)) in the simulation.  These demonstrate that both species are accelerated by Fermi acceleration into power-laws, consistent with the highly correlated ion and electron acceleration observed in solar flares \citep{Shih2009}.

\textit{Discussion.---}
While observations have suggested efficient acceleration of both electrons and ions during nonrelativistic reconnection in solar flares and the magnetotail, establishing this from first-principle kinetic simulations has been a long-standing challenge. 
For the first time, our simulations produce power-law distributions for both electrons and protons that contain a significant amount of released energy, providing a plausible explanation to the solar flare observations \cite{Lin2011,Omodei2018,Gary2018,Krucker2010,Oka2015,Aschwanden_2016}. The $p \sim 4$ spectra obtained in our simulations are consistent with the electron indices inferred from many x-ray and microwave observations \citep{Oka2018,Gary2018}, and the proton indices from gamma-ray \citep{Omodei2018} and SEP observations \citep{Cohen2020}. For the September 10, 2017 event observed by numerous instruments, comparison between MHD simulations and gyrosynchrotron emission suggests reconnection occurred with a weak guide field ($b_g \sim 0.3$) \citep{Chen2020}. Evidence of turbulent reconnection has been presented \citep{Cheng2018,French2019} for this event and the power-law index obtained by gyrosynchrotron emssion is $p\sim 3.5-6.5$, broadly consistent with our simulations.  
Our simulations can also be compared positively with a well-observed magnetotail event reported by Magnetospheric Multi-scale Mission \citep{Ergun2018,Ergun2020}. The event shows simultaneous electron and proton nonthermal acceleration in a broad turbulent reconnection region over $\sim 16$ Earth radii ($\sim 80 d_i$) with a low upstream $\beta$ ($\beta_e\sim0.03$), quite similar to our simulations. The observed power-law indices are typically $\sim 3.3-4.3$ for protons and $\sim 4.2-5.4$ for electrons, in agreement with our simulations. The shoulders of the observed spectra are $\sim 15$ keV ($0.2 m_i V_A^2$) for electrons and $\sim 40$ keV ($0.6 m_iV_A^2$) for protons, also similar to our simulations. Moreover, protons are observed to gain more energy than electrons.

We have demonstrated that flux-rope kink instability drives strong field-line chaos in 3D reconnection with weak guide fields, leading to strong particle transport and acceleration. As a result, both electrons and protons are accelerated into clear power-laws, whose basic properties such as efficiency and spectral indices are controlled by the injection and Fermi acceleration processes. The formation of the power-laws, especially protons, requires large domain size in the reconnection plane and long acceleration time, as well as sufficient 3D physics to capture the flux-rope $m=1$ kink instability.
This work uncovers the fundamental processes for initializing and developing nonthermal ion and electron acceleration in nonrelativistic magnetic reconnection, with strong implications to not only heliophysics but also astrophysics such as stellar flares and accretion-disk flares \citep{Ripperda2020,Nathanail2020}. 

\textit{Acknowledgment.---}
We gratefully acknowledge the helpful discussions in the SolFER DRIVE Science Center collaboration. We also acknowledge technical support from Bin Dong, Suren Byna, and K. John Wu at Lawrence Berkeley National Laboratory. Q.Z., F.G., W.D., and H.L. acknowledge the support from the U.S. Department of Energy, Office of Science, Office of Fusion Energy Sciences, and from Los Alamos National Laboratory, through the LDRD
program and its Center for Space and Earth Science
(CSES), and from NASA programs through Grants No.
NNH17AE68I, No. 80HQTR20T0073, No. 80NSSC20
K0627, and No. 80HQTR21T0005, and through the
Astrophysical Theory Program. The work by X.L. is
funded by the National Science Foundation Grant
No. PHY-1902867 through the NSF/DOE Partnership in
Basic Plasma Science and Engineering and NASA MMS
80NSSC18K0289. The simulations used resources provided by the Los Alamos National Laboratory Institutional
Computing Program (supported by the U.S. Department of
Energy National Nuclear Security Administration), the
National Energy Research Scientific Computing Center
(NERSC, a U.S. Department of Energy Office of Science
User Facility at Lawrence Berkeley National Laboratory),
and the Texas Advanced Computing Center (TACC, at the
University of Texas at Austin).

\section{Supplemental Material}
Table \ref{table1} lists information of all simulations discussed in this paper with a broad range of parameters. Using a large set of fully kinetic simulations (Runs 1-15), we verified the threshold $L_{th}\sim \pi b_g D\sim 0.1\pi b_g L_x$ for the development of flux-rope kink instability and the field-line chaos, which is a necessary condition for the power-law formation. Runs 1-11 are used to verify the dependence on domain size, while additional Runs 12-15 are for verifying the dependence on guide fields in the weak-guide-field regime. With Run 16, we also confirmed the low-energy bounds of proton power-laws $\varepsilon_{li}\sim0.5 m_i V_A^2$ as long as $\beta \lesssim 0.1$. With Runs 17-18, we also verified that the electron and proton spectra are insensitive to mass ratios. 
\begin{widetext}
\begin{center}
\begin{table}
\renewcommand\thetable{S1} 
\begin{tabular}{  c c c c c c c c c c  } 
 \hline
Run	& $L_x/d_i$	& $L_y/d_i$	& $L_z/d_i$ &   $\beta_e$ &$B_g/B_0$&$m_i/m_e$&$L_{th}/d_i$& \makecell{m=1 \\ kink unstable}\\
 \hline
1	& 300	& 25.0 & 125.0 & 0.02 &0.2&25& 19&	Yes \\
2	& 300	& 12.5 & 125.0 & 0.02 &0.2&25&  19&	 No\\
3	& 150	& 25.0 & 62.5 & 	0.02 &0.2&25&  9.5&	Yes \\
4	& 150	& 12.5 & 62.5 & 	0.02 &0.2&25& 9.5& 	Yes \\
5	& 150	& 6.25 & 	62.5 & 	0.02 &0.2&25& 9.5& 	 No \\
6	& 150	& 3.125 & 	62.5 & 	0.02 &0.2&25& 9.5& 	 No \\
7	& 150	& 1.5625 & 	62.5 & 	0.02 &0.2&25&  9.5&	 No \\
8	& 150	& 0.7813 & 	62.5 & 	0.02 &0.2&25&  9.5&	 No \\
9	& 150	& 0.3906 & 	62.5 & 	0.02 &0.2&25& 9.5& 	 No \\
10	& 150	& 0.1465 & 	62.5 & 	0.02 &0.2&25& 9.5& 	 No \\
11	& 150	& 0.0488 & 	62.5 & 	0.02 &0.2&25& 9.5& 	 No \\
12	& 150	& 18.75 & 62.5 & 	0.02 &0.3&25& 14.25& 	Yes \\
13	& 150	& 9.375 & 62.5 & 	0.02 &0.3&25& 14.25&	No \\
14	& 150	& 7.8125 & 62.5 & 	0.02 &0.1&25& 4.75&	Yes \\
15	& 150	& 3.125 & 	62.5 & 	0.02 &0.1&25& 4.75&	 No \\
16	& 150	& 12.5 & 62.5 & 	0.08 &0.2&25& 9.5&	Yes \\
17	& 75	& 6.25 & 31.25 & 	0.02 &0.2&25& 4.75&	Yes\\
18	& 75	& 6.25 & 31.25 & 	0.02 &0.2&100&4.75& 	Yes\\
\hline
\end{tabular}
\caption{All simulations discussed in this paper.\label{table1}}
\end{table}
\end{center}
\end{widetext}

%


\end{document}